\def\dir{./}
\documentclass[useAMS,usenatbib]{\dir mn2e}
\usepackage[fleqn]{amsmath}
\usepackage{amssymb}
\usepackage[super]{nth}
\usepackage{natbib}
\usepackage{graphicx}
\usepackage{float}
\usepackage{mathrsfs}
\usepackage{mathtools}
\usepackage{multirow}
\usepackage[usenames,dvipsnames]{color}
\usepackage{\dir aas_macros}
\usepackage{comment}
\usepackage{subfigure}
\usepackage{hyperref}
\usepackage{mathrsfs}
\usepackage{epstopdf, dcolumn}
\DeclareGraphicsExtensions{.pdf}
\usepackage{wasysym}
\usepackage{dashrule}
\usepackage{xcolor}
\usepackage{arydshln}
\usepackage{threeparttable}
\usepackage[export]{adjustbox}
\usepackage[figuresleft]{rotating}

\newlength\replength
\newcommand\repfrac{.33}

\newcommand\rulewidth{.6pt}
\newcommand\tdashfill[1][\repfrac]{\cleaders\hbox to \replength{%
		\smash{\rule[\arraystretch\ht\strutbox]{\repfrac\replength}{\rulewidth}}}\hfill}

\newcommand\tdotfill[1][\repfrac]{\cleaders\hbox to \replength{%
		\smash{\raisebox{\arraystretch\dimexpr\ht\strutbox-.1ex\relax}{.}}}\hfill}

\newcommand{\appropto}{\mathrel{\vcentre{
			\offinterlineskip\halign{\hfil$##$\cr
				\propto\cr\noalign{\kern2pt}\sim\cr\noalign{\kern-2pt}}}}}

\newcommand{\meraxes}{{\textsc{\small Meraxes}}}
\newcommand\lsim{\mathrel{\rlap{\lower4pt\hbox{\hskip1pt$\sim$}}
        \raise1pt\hbox{$<$}}}
\newcommand\gsim{\mathrel{\rlap{\lower4pt\hbox{\hskip1pt$\sim$}}
        \raise1pt\hbox{$>$}}}
\newcommand{\Rom}[1]{\uppercase\expandafter{\romannumeral #1}}
\newcommand{\rom}[1]{\lowercase\expandafter{\romannumeral #1}}

\newcommand{\hone}{\mathrm{H}\textsc{i}}
\newcommand{\hii}{\mathrm{H}\textsc{ii}}
\newcommand{\tocm}{{$\mathrm{21cm}\textsc{fast}$}}
\newcommand{\lya}{{Ly$\alpha$}}

\newcommand{\msol}{{\rm M}_\odot}

\begin{document}

\title[Damping-wing optical depth]{\mbox{Dark-ages Reionization and Galaxy Formation Simulation XX.} {The {\lya} IGM transmission properties and environment of bright galaxies during the Epoch of Reionization}}

\author[Qin et al.]{Yuxiang Qin$^{1,2,3}$\thanks{E-mail: Yuxiang.L.Qin@gmail.com}, J. Stuart B. Wyithe$^{1,2}$, Pascal A. Oesch$^{4,5}$, Garth D. Illingworth$^6$, 
	\newauthor Ecaterina Leonova$^{4}$, Simon J. Mutch$^{1,2}$ and Rohan P. Naidu$^{7}$ \\
	$^{1}$School of Physics, University of Melbourne, Parkville, VIC 3010, Australia\\
	$^{2}$ARC Centre of Excellence for All Sky Astrophysics in 3 Dimensions (ASTRO 3D)\\
	$^{3}$Scuola Normale Superiore, Piazza dei Cavalieri 7, I-56126 Pisa, Italy\\
	$^{4}$Department of Astronomy, University of Geneva, Chemin Pegasi 51, 1290 Versoix, Switzerland\\
	$^{5}$Cosmic Dawn Center (DAWN), Niels Bohr Institute, University of Copenhagen, Jagtvej 128, K\o benhavn N, DK-2200, Denmark\\
	$^{6}$Department of Astronomy and Astrophysics, UCO/Lick Observatory, University of California, Santa Cruz, CA 95064, USA\\
	$^{7}$Center for Astrophysics $|$ Harvard \& Smithsonian, 60 Garden Street, Cambridge, MA 02138, USA
}
\maketitle
\label{firstpage}

\begin{abstract}
The highly neutral inter-galactic medium (IGM) during the Epoch of Reionization (EoR) is expected to suppress {\lya} emission with damping-wing absorption, causing nearly no {\lya} detection from star-forming galaxies at $z{\sim}8$. However, spectroscopic observations of the 4 brightest galaxies (${\rm H}_{160}{\sim}25$ mag) at these redshifts do reveal prominent {\lya} line, suggesting locally ionised IGM. In this paper, we explore the {\lya} IGM transmission and environment of bright galaxies during the EoR using the Meraxes semi-analytic model. We find brighter galaxies to be less affected by damping-wing absorption as they are effective at ionizing surrounding neutral hydrogen. Specifically, the brightest sources (${\rm H}_{160}{\lesssim}25.5$ mag) lie in the largest ionized regions in our simulation, and have low attenuation of their {\lya} from the IGM (optical depth ${<}1$). Fainter galaxies (25.5 mag${<}{\rm H}_{160}{<}27.5$ mag) have transmission that depends on UV luminosity, leading to a lower incidence of {\lya} detection at fainter magnitudes. This luminosity-dependent attenuation explains why {\lya} has only been observed in the brightest galaxies at $z{\sim}8$. Follow-up observations have revealed counterparts in the vicinity of these confirmed $z{\sim}8$ {\lya} emitters. The environments of our modelled analogues agree with these observations in the number of nearby galaxies, which is a good indicator of whether {\lya} can be detected among fainter galaxies. At the current observational limit, galaxies with ${\ge}2$--5 neighbours within $2'{\times}2'$ are ${\sim}2$--3 times more likely to show {\lya} emission. {\it JWST} will discover an order of magnitude more neighbours, revealing ${\gtrsim}50$ galaxies in the largest ionizing bubbles and facilitating direct study of reionization morphology.
\end{abstract}


\section{Introduction}
The transmission of Lyman-$\alpha$ ({\lya}) photons has played a significant role in observational studies of reionization due to its highly resonant nature in neutral hydrogen, which provides a direct probe of the ionization state of the intergalactic medium (IGM). Toward the end of reionization, the large-scale $\hone$ distribution has been studied with the {\lya} forest along the line of sight towards bright background quasars  \citep{Bolton2010MNRAS.406..612B,Lidz2010ApJ...718..199L,Lee2015ApJ...799..196L,Puchwein2015MNRAS.450.4081P,Bolton2017MNRAS.464..897B,Gaikwad2020MNRAS.494.5091G}. For instance, by counting the zero-flux pixels on quasar spectra, \citet{McGreer2015MNRAS.447..499M} concludes that the average neutral hydrogen fraction ($x_{\hone}$) at $z\sim5.9$ should be lower than 6 -- 11 per cent ($1\sigma$; see also upcoming works by Campo et al., in prep. and Jin et al., in prep.). However, individual sight-lines seem to suggest that large chunks (${>}160$cMpc) of residual $\hone$ still exist even at $z<6$ (\citealt{Becker2015MNRAS.447.3402B}; Zhu et al., in prep.), requiring late reionization and/or that the underlying accountable sources possess unusual ionizing properties (e.g. \citealt{DAloisio2015ApJ...813L..38D,DAloisio2018MNRAS.473..560D,Chardin2015MNRAS.453.2943C,Chardin2017MNRAS.465.3429C,Davies2016MNRAS.460.1328D,Keating2018MNRAS.477.5501K,Keating2020MNRAS.491.1736K,Kulkarni2019MNRAS.485L..24K,Meiksin2020MNRAS.491.4884M,Nasir2020MNRAS.494.3080N,Qin2021arXiv210109033Q}).

Damping-wing absorption of {\lya} emission lines of quasars (in particular the red side of the line center on a quasar spectrum) also facilitates estimates of $x_{\hone}$ during the epoch of reionization (EoR). This is done by directly comparing the observed flux to the intrinsic flux inferred from low-redshift analogues using continuum and metal lines. For instance, among some of the 9 quasars discovered at $z{>}7$ to date (\citealt{Bosman2020zndo...3634964B} and reference therein), high-quality spectra have enabled direct measurement of the neutral hydrogen, ranging from $x_{\hone}{\sim}10$\% to 80\%  \citep{Mortlock2011Natur.474..616M,Greig2017MNRAS.466.4239G,Greig2019MNRAS.484.5094G,Banados2018Natur.553..473B,Davies2018ApJ...864..142D,Wang2020ApJ...896...23W,Yang2020ApJ...897L..14Y}.

High-redshift star-forming galaxies that have detectable {\lya} emission (e.g. with a equivalent width larger than $25{\rm \AA}$) are referred to as {\lya} emitting galaxies (LAEs). The incidence of LAEs has also been used to determine the neutral fraction \citep{Mason2018ApJ...857L..11M,Mason2018ApJ...856....2M}. This is because {\lya} detection during the EoR could provide evidence for whether galaxies sit in large ionized bubbles, so that their {\lya} photons are redshifted deep into the damping wing, and are therefore less attenuated by the intergalactic ${\hone}$ \citep{Dijkstra2014PASA...31...40D}. As the Universe becomes more neutral, observations have shown a rapidly decreasing fraction of LAEs among Lyman break galaxies (LBGs) from $z=6$ to 8 \citep{Fontana2010ApJ...725L.205F,Stark2010MNRAS.408.1628S,Ono2012ApJ...744...83O,Treu2013ApJ...775L..29T,Schenker2014ApJ...795...20S,Hoag2019ApJ...878...12H}. 

However, despite the increasing neutral fraction in the high-redshift Universe, prominent {\lya} lines in some of the brightest sources have been observed even out to $z{\sim}9$ with a much higher detection rate of {\lya} emission than expected. In particular, the 4 most luminous galaxies ($H_{160}{\sim}25$ mag) in CANDELS have all been spectroscopically confirmed by Keck as LAEs ( \citealt{Oesch2015ApJ...804L..30O,Zitrin2015ApJ...810L..12Z,Roberts-Borsani2016ApJ...823..143R,Stark2017MNRAS.464..469S}; see also \citealt{Tilvi2020ApJ...891L..10T}). In contrast, when targeting relatively fainter galaxies, no {\lya} emission was detected at $z{\sim}8$ while $p({\rm Ly}\alpha)$ was found to be ${<}$2\% and ${<}$15\% at $z{\sim}7.5$ and 7, respectively \citep{Schenker2014ApJ...795...20S,Hoag2019ApJ...878...12H}. Similarly, more recent larger spectroscopic samples of {\lya} emission among high-redshift, bright galaxies find $p({\rm Ly}\alpha)$ varying between 20\% and 50\% at $6{\lesssim}z{\leq}9$ \citep{Jung2020arXiv200910092J,Endsley2020arXiv201003566E,Laporte2021arXiv210408168L}. Apart from possibly having higher intrinsic {\lya} emission due to recent star formation, these brightest galaxies are likely located in more overdense regions, and may be surrounded by a larger number of star-forming galaxies producing a more transparent IGM (\citealt{Wyithe2005ApJ...625....1W}).  

In this work, we use a semi-analytic galaxy-formation model to explore the {\lya} IGM transmission properties and environment of bright galaxies during the EoR. We focus on both the brightest galaxies, which are analogous to those observed at $z{\sim}8$, as well as relatively fainter galaxies, including those that can be considered promising candidates for the {\it James Webb Space Telescope} ({\it JWST}; e.g. \citealt{Malkan2021jwst.prop.1571M,Naidu2021jwst.prop.2279N,Oesch2021jwst.prop.1895O}). Our semi-analytic model (SAM) follows a number of astrophysical processes of galaxy formation and self-consistently calculates photon-heating feedback from reionzation as well as recombination in $\hii$ regions. The model has been shown to successfully reproduce a large number of observables \citep{Mutch2016MNRAS.463.3556M,Liu2016,Qin2017a} including galaxy \citep{Bouwens2015a,Bouwens2016} and quasar UV luminosity functions (LFs; \citealt{wolf2003,Bongiorno2007vvds,Croom2009,Glikman2011,Masters2012,Palanque-Delabrouille13}) in addition to the inferred EoR history from the comic microwave background (CMB;  \citealt{Planck2020A&A...641A...6P}, see also \citealt{Qin2020arXiv200616828Q}).% These are crucial for further investigation of the LAE population within a cosmological context.

This paper is organized as follows. We provide a brief review of our model as well as the newly implemented calculation of {\lya} damping-wing absorption in Section \ref{sec:model}. We present our results including the strategy for hunting LAEs in Section \ref{sec:result} before concluding in Section \ref{sec:conclusion}. In this work, we set the following cosmological parameters based on the {\it Planck} 2015 results ($\Omega_{\mathrm{m}}, \Omega_{\mathrm{b}}, \Omega_{\mathrm{\Lambda}}, h, \sigma_8, n_s $ = 0.308, 0.0484, 0.692, 0.678, 0.815, 0.968; \citealt{Planck2016A&A...594A..13P}).

\section{Modeling high-redshift galaxy properties and reionizaton}\label{sec:model}
We use the {\meraxes} SAM \citep{Mutch2016,Qin2017a} from the DRAGONS (Dark-ages Reionization And Galaxy formation Observables from Numerical Simulations) program, which is coupled to a parent {\it N}-body simulation \citep{Poole2016MNRAS.459.3025P} providing dark matter halo properties from $z=35$ to 2. The upper left panel of Fig. \ref{fig:snapshot} illustrates the matter density at $z=8$ in our simulation volume. With a mass resolution of  ${\sim}4{\times}10^6{\msol}$ and a box size of 100cMpc, we are able to properly sample low-mass halos down to the atomic cooling threshold and capture massive galaxies of ${\sim}7{\times}10^{11}{\msol}$ at $z=8$. These allow us to resolve faint galaxies that are responsible for reionization \citep{Liu2016} and study galaxies as bright as $M_{1600}{\sim}-22$ mag at $z=11$ \citep{Mutch2016MNRAS.463.3556M} or ${\rm H}_{160}{=}24.5$ mag (i.e. the {\it Hubble Space Telescope} or {\it HST} F160W band) at $z=8$. While we refer interested readers for detailed semi-analytic prescriptions of galaxy formation to \citet{Mutch2016} and \citet{Qin2017a}, some basic characteristics of our model are presented below.

\subsection{High-redshift galaxies}

\begin{figure*}
	\begin{minipage}{\textwidth}
		\centering
		\includegraphics[width=\textwidth]{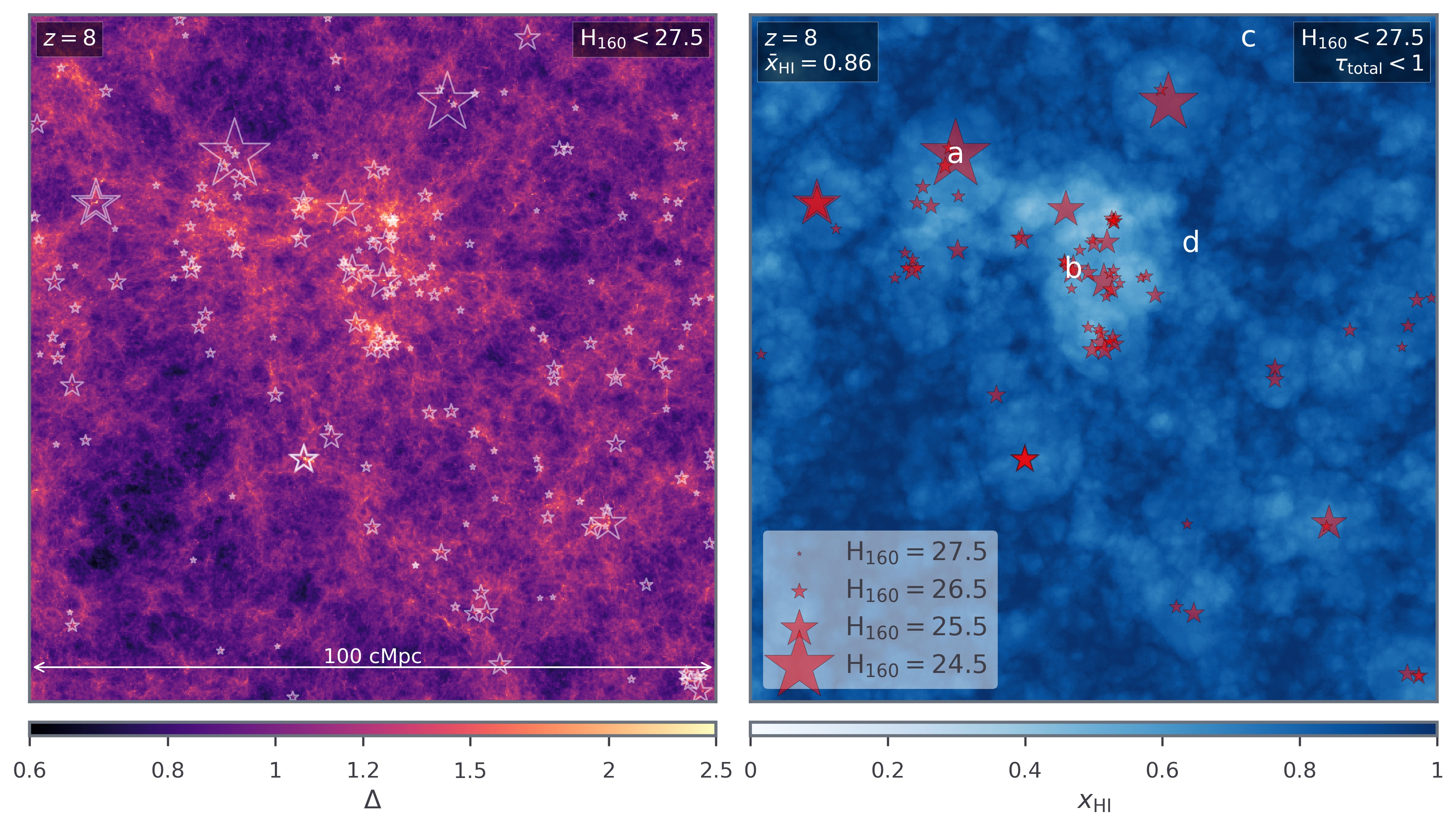}\\%\vspace*{-2mm}\\
		\includegraphics[width=\textwidth]{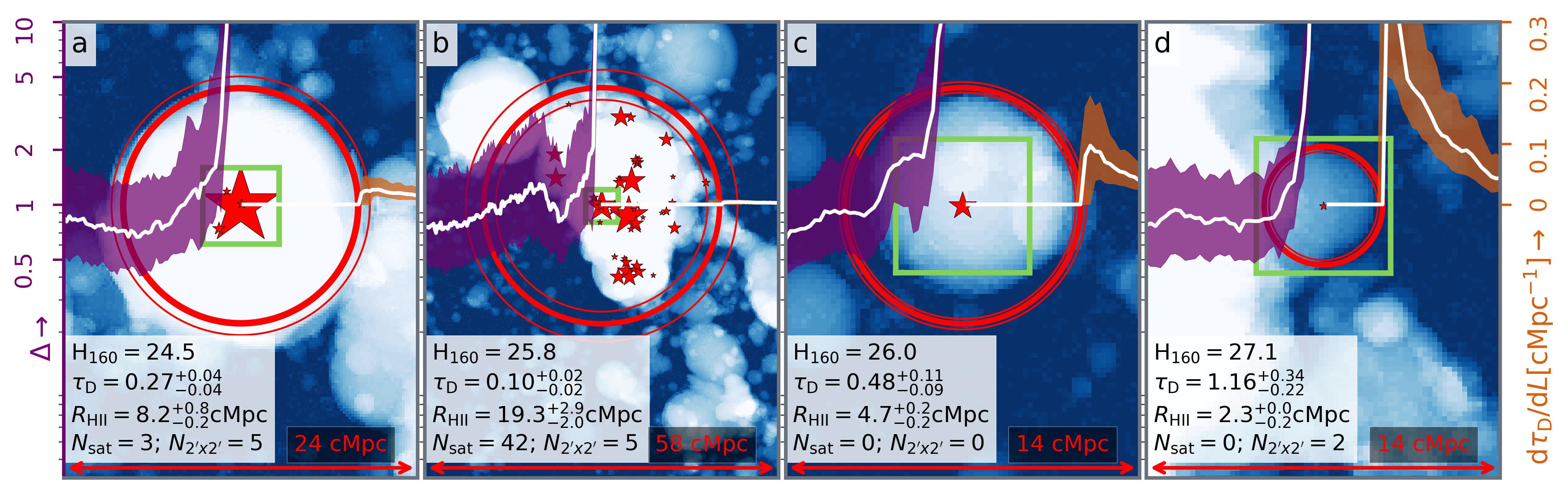}%\vspace*{-2.5mm}
		\caption{\label{fig:snapshot}
			{\it Upper panels:} overdensity ($\Delta$) and neutral hydrogen fraction ($x_{\hone}$) of the entire simulation volume at $z{=}8$ (projected with a depth of 100cMpc). In the density plot, galaxies brighter than ${\rm H}_{160}{=}27.5$ mag are indicated using star symbols with increasing sizes representing higher luminosities. In the $\hone$ plot, only UV-bright, {\lya}-transparent (i.e. $\tau_{\rm total}{<}1$) galaxies are shown and the location of four example galaxies (a, b, c, d; see more in Section \ref{subsubsec:showcases}) are indicated.
			{\it Bottom panels:} highlights of four bright galaxies (zoom-in projection with a depth of 10cMpc for $x_{\hone}$ and bright galaxies within the same $\hii$ bubble). 
			{Radial profiles of $\Delta$ (purple; left-hand side) and the damping-wing optical depth gradient (${\rm d}\tau_{\rm D}/{\rm d}L$; yellow; right-hand side) are shown with their medians and uncertainties ([16, 84] percentiles) among 500 random sight-lines highlighted with white solid curves and colored regions, respectively. 
			Their vertical ranges are presented on the left- and right-hand side of the bottom panels, separately.}
			%The ionizing bubble is indicated by the thick red circle (and thin circles for uncertainties). 
			Presented in the lower left corner of each sub-panel are the target galaxy luminosity, damping-wing optical depth and bubble radius, followed by the number of neighbouring bright galaxies (${\rm H}_{160}{<}27.5$ mag) within the $\hii$ bubble (mean radius indicated by thick red circles with the thin ones for uncertainties) and that within a {\it HST} WFC3 FoV ($2'{\times}2'$'; 2D projection with $\Delta z{\sim}1$; green square).% See the text in Section \ref{subsubsec:showcases} for more details.
		}
	\end{minipage}
\end{figure*}

{\meraxes} evaluates the accretion and cooling of gas, stellar evolution and feedback, growth of supermassive blackholes and their impact on galaxy formation, as well as other environmental influences such as mergers and photo-heating from reionization. We also consider dust attenuation and forward model the observed galaxy and quasar UV magnitudes (e.g. ${\rm H}_{160}$). These allow us to adjust our model parameters in order to match the predicted observables with existing measurements during the EoR. These include the stellar mass function, UV LFs of LBGs and high-redshift quasars, ionizing emissivities, and the Thomson scattering optical depth of the CMB photons. In this work, we use the galaxy catalogue generated by the fiducial model presented in \citet{Qin2017a}. As an illustration, Fig. \ref{fig:snapshot} populates the modelled bright galaxies (${\rm H}_{160}{<}27.5$ mag; i.e. the current CANDELS DEEP $5\sigma$ depth, see e.g. \citealt{Roberts-Borsani2016ApJ...823..143R}) in the density plot with larger star symbols indicating higher luminosities.

Modelling the intrinsic {\lya} profile of individual galaxies \citep{Verhamme2006A&A...460..397V,LeDelliou2006MNRAS.365..712L,Dijkstra2007MNRAS.377.1175D,Zheng2010ApJ...716..574Z,Song2020ApJ...901...41S,Garel2021arXiv210403339G} is deferred to future work. Here, we assume all star-forming galaxies during the EoR are effective at producing {\lya} photons and their detectability in {\lya} is solely determined by neutral hydrogen present along the line of sight (e.g. \citealt{Gronke2020arXiv200414496G}; see more discussion in Section \ref{subsec:lyalpha}).

\subsection{Ionizing the IGM}
To evaluate the IGM ionization state, {\meraxes} couples its UV ionizing photon source (i.e. star-forming galaxies and quasars) model with {\tocm} \citep{Mesinger2011MNRAS.411..955M,Murray2020JOSS....5.2582M}. We divide the simulation volume into $N{=}512^3$ cells having an equal length of $L{=}0.195$cMpc. With an excursion-set algorithm \citep{Furlanetto2004ApJ...613....1F}, we compare the ionizing photon budget to the number of present neutral hydrogen atoms (including those recombined after ionization) and estimate the neutral hydrogen fraction ($x_{\hone}$) in each cell.

The model used in this work has a predicted EoR history consistent with the latest {\it Planck} results \citep{Planck2016A&A...594A..13P,Planck2020A&A...641A...6P,Qin2020arXiv200616828Q}. It is predominately driven by UV ionizing photons emitted by stars in faint galaxies \citep{Liu2016} with supermassive blackholes having an insignificant contribution to the overall reionization (e.g. ${\lesssim}5$ per cent of the ionizing production rate at $z\ge7$; \citealt{Qin2017a}). The only regions of ionizing background dominated by luminous AGN are in their immediate environment \citep{Qin2017MNRAS.471.4345Q}. The upper right panel of Fig. \ref{fig:snapshot} illustrates the modelled ionization field at $z=8$ where the volume-averaged neutral hydrogen fraction is $\bar{x}_{\hone}=0.86$. We see reionization (anti-)correlates with underlying densities and brighter galaxies, in general, are located in larger $\hii$ bubbles. We also confirm that, in these star-forming galaxies, the central supermassive black holes have a negligible role in ionizing the surrounding IGM.

\subsection{{\lya} damping-wing absorption}\label{subsec:lyalpha}

We compute the {\lya} optical depth in each cell based on
\begin{equation}\label{eq:tau_lya}
	\tau_{\alpha} =L n_{\rm H,0} x_{\hone}  \Delta (1+z)^2 \sigma_{\alpha},
\end{equation}
where $L$, $n_{\rm H,0}$ and $\Delta$ are the pixel length, the cosmic mean of the comoving hydrogen number density and its local overdensity in each cell, while
\begin{equation}\label{eq:sigma_lya}
	\sigma_{\alpha}\left(x{\equiv}\frac{\lambda_\alpha}{\lambda}\right) = \frac{3\lambda_{\alpha}^2}{8\pi} \frac{x^4}
	{1/4x^6 + (1-x)^2({2\pi c}/{\lambda_\alpha\Lambda})^2}
\end{equation}	
corresponds to the {\lya} scattering cross section in {$\hone$} redwards of the line centre (i.e. $x{<}1$; \citealt{MiraldaEscide1998ApJ...501...15M}) with c and $\Lambda=6.25\times10^8 {\rm s}^{-1}$ representing the speed of light and the decay constant for the {\lya} resonance. From these two equations, we see that damping-wing absorption increases towards higher redshifts [$\tau_{\alpha}{\propto} (1{+}z)^2$] and drops significantly ($\sigma_{\alpha}{\propto} \lambda^{-4}$; i.e. Rayleigh-type scattering) at distances further from the line center [i.e. $x\ll 1{-}\Lambda\lambda_{\alpha} /(2\pi c)$]. Therefore, we ignore damping-wing absorption at distances longer than our box length (i.e. 100cMpc) and estimate the total optical depth as $\tau_{\rm D} =  \sum_{i=0}^{\sqrt[3]{N}-1} \tau_{\alpha,i} $. Here, $\tau_{\alpha,i} $ represents the optical depth in the $i$th cell away from the source along a sight-line where {\lya} has shifted to ${\lambda}{=}{\lambda_\alpha}\left[1{+}\left(V_{\rm offset}+iLH\right)/c\right]$ with $V_{\rm offset}$ and $H(z)$ representing the velocity offset of the intrinsic {\lya} line and the Hubble parameter at $z$. Note that, for each bright galaxy, we randomly select 500 sight-lines and estimate the bubble size\footnote{For each target galaxy, we consider its distance to the first cell on the sight-line with an ionization phase transition (i.e. $x_{\hone}{\ge}0.9$ in this work) as its bubble radius.} and $\tau_{\rm D}$ as well as their associated uncertainties by taking the median and [16, 84] percentiles.

There are an increasing number of observed LAEs presenting a velocity offset between the {\lya} emission peak and the systemic redshift up to $V_{\rm offset}{\sim}1000{\rm km\ s^{-1}}$, as well as a possible correlation (using large samples at low redshifts) showing higher $V_{\rm offset}$ with increasing galaxy UV luminosities (\citealt{Erb2014ApJ...795...33E,Muzahid2020MNRAS.496.1013M}; Endsley et al., in prep.). This could increase the {\lya} transmission in the IGM for brighter galaxies \citep{Dijkstra2007MNRAS.377.1175D,Choudhury2015MNRAS.452..261C,Mason2018ApJ...857L..11M,Mason2018ApJ...856....2M}. However, when splitting low-redshift LAEs by their UV magnitudes, we see the opposite with more luminous galaxies showing a lower fraction of LAEs (e.g. \citealt{Schenker2014ApJ...795...20S,Garel2015MNRAS.450.1279G}), possibly due to a reduced {\lya} escape fraction in deeper gravitational potentials \citep{Oyarzun2017ApJ...843..133O,Yang_2017,Hassan2021ApJ...908..219H}. As current measurements are still inconclusive and limited by the small LAE sample, particularly at higher redshifts, we do not consider the intrinsic line intensity, offset or escape fraction of {\lya}. Since the targets in this work are those mostly sitting in large {$\hii$} bubbles, we also expect them to be less affected by $V_{\rm offset}$. Based on these, we treat the detectability of a LAE governed primarily by its IGM transmission.

It is worth noting that fully ionized regions do not contribute to the total damping-wing optical depth according to equation (\ref{eq:tau_lya}) as $x_{\hone}{=}0$. However, to account for resonant absorption by residual neutral hydrogen inside the ionized region \citep{Mesinger2015MNRAS.446..566M,Park2021arXiv210510770P}, we follow previous work (e.g. \citealt{Mason2018ApJ...856....2M}) and assume all flux bluer than the circular velocity of the host halo becomes attenuated. This leads to a factor of 2 further reduction to the {\lya} transmission ($\mathcal{T}$) so that
\begin{equation}\label{eq:T}
	\mathcal{T} \equiv \exp(-\tau_{\rm total}) \equiv \exp(-\tau_{\rm D}-\tau_{\rm HII}) = 0.5\exp(-\tau_{\rm D}).
\end{equation}
This results in an optical depth inside {$\hii$} regions of $\tau_{\rm HII}{\sim}0.69$ while $\tau_{\rm total}$ in equation (\ref{eq:T}) represents the total optical depth.

In this work, we consider {\it galaxies with $\tau_{\rm total}$ larger (or lower) than 1 as highly attenuated (or transparent) LAEs subject to damping-wing absorption in the IGM}. The upper right panel of Fig. \ref{fig:snapshot} shows the location of UV-bright, transparent-LAEs. We see a large fraction (${>}60$\%) of UV-bright LBGs are excluded (c.f. the upper left panel of Fig. \ref{fig:snapshot}) as they are still too faint and/or isolated from other bright sources to ionize their surrounding neutral hydrogen. 

\subsection{Example LAEs}\label{subsubsec:showcases}

We highlight four example galaxies in the bottom panels of Fig. \ref{fig:snapshot}. These galaxies are, in each case, the brightest galaxy found in their $\hii$ bubble. Hence, we also refer to them as the central galaxy and consider the remaining counterparts in the bubble as their satellites\footnote{As some bubbles are highly non-spherical, e.g. Galaxy b in Fig. \ref{fig:snapshot}, there are scenarios where a brighter galaxy, the central of its own $\hii$ bubble ($R_{\hii,{\rm bright}}$), can be considered as a satellite of an other galaxy which is fainter but has a bubble size ($R_{\hii,{\rm faint}}$) larger than the brighter one. Note that this fainter galaxy is not categorised as a satellite of the brighter one because their separation is longer than $R_{\hii,{\rm bright}}$ but shorter than $R_{\hii,{\rm fainter}}$}. In each sub-panel, we also illustrate the overdensity (purple; left-hand side) and {\lya} damping-wing optical depth {as in ${\rm d}\tau_{\rm D}/{\rm d}L$} (yellow; right-hand side) among 500 random sight-lines towards the target galaxy.

% in cells along the 500 random sight-lines towards the target galaxy with the median and uncertainties corresponding to [16, 84] percentiles highlighted with white solid curves and colored shaded regions. Note that the vertical range from the target galaxy position to the upper boundary of the subpanel (e.g. the black vertical line with arrow in panel c) represents $\Delta=1$--10 in a logarithmic scale and $\tau_{\alpha}=0$--0.05 in a linear one, respectively. The ionizing bubble is indicated by the thick red circle (and thin circles for uncertainties), with the number of satellites within the median bubble size shown in the lower left corner, following the galaxy luminosity, damping-wing optical depth and bubble radius. 

Furthermore, we consider {\it HST} WFC3 mock images centred at the target galaxies (see the green square; same projected direction as {$\hii$}) and count the number of neighbouring ${\rm H}_{160}{<}27.5$ mag galaxies within a field of view (FoV) of $2'{\times}2'$ and a depth of $\Delta z{\sim}1$. The latter is a conservative choice set by the photometric uncertainty in relevant observations ($\Delta z{\sim}{\pm}0.1$--0.4, e.g. \citealt{Roberts-Borsani2016ApJ...823..143R}; see also \citealt{Bouwens2015a}). Note that as our box length sets the observational depth to be $\Delta z{\sim}0.35$ at $z{=}8$, ${\sim}3$ times smaller than the desired value, we extend each mock observation with two more randomly selected sight-lines before projection. To account for galaxy evolution, we choose the sight-lines from simulation snapshots at $z{=}7.5$ and 8.5 separately. These allow us to account for foreground and background galaxies within our FoV due to random alignment, the number of which is $1\pm1$ (median with [16, 84] percentiles). Below, we discuss each of the 4 example galaxies shown in Fig. \ref{fig:snapshot} in turn.
\begin{enumerate}
	\item[\textbf{Galaxy a}.] The brightest galaxy in our simulation snapshot with ${\rm H}_{160}{=}24.5$ mag. This galaxy is considered to be an observable LAE with the UV ionizing photons in this area coming predominately from itself and resulting in a {\lya} damping-wing optical depth of $\tau_{\rm D}{\sim}0.27$. There are also 3 satellite galaxies with ${\rm H}_{160}{<}27.5$ mag in its $\hii$ bubble with a radius of $R_{\rm HII}{\sim}8$cMpc. In the mock observation, we expect to find 2 more galaxies randomly aligned in the FoV in addition to the 3 satellites.
	\item[\textbf{Galaxy b}.] A luminous galaxy (${\rm H}_{160}{=}25.8$ mag) located in one of the largest $\hii$ regions in our simulation with $\tau_{\rm D}{\sim}0.10$. Centred at this particular galaxy, we obtain a highly non-spherical $\hii$ bubble (i.e. $R_{\hii}{\sim}19$cMpc with a large uncertainty of $2-3$cMpc). Within a sphere of $R_{\hii}{\sim}19$cMpc, more than 40 galaxies with ${\rm H}_{160}{<}27.5$ mag contribute to the local ionizing photon budget and 5 of them are expected in the mock observation (no random alignment). This galaxy has a damping-wing optical depth much smaller than the brightest one (i.e. Galaxy a), suggesting that bright galaxies are more likely to have observable {\lya} emission when clustered (\citealt{Wyithe2005ApJ...625....1W,Trac2011ASL.....4..228T,Endsley2020arXiv201003566E}; see also the recent report on a LAE cluster at $z{\sim}7$ by \citealt{Hu2021NatAs...5..485H} and an upcoming result from Jung et al., in prep. showing boosted IGM transmission around bright galaxies at $z{>}6$). 
	\item[\textbf{Galaxy c}.] The brightest galaxy in our simulation that is considered to be an obscured LAE. This galaxy has a similar UV magnitude (${\rm H}_{160}{=}26$ mag) to Galaxy b, but is located in a much smaller $\hii$ bubble (${\sim}5$cMpc) and possesses a $\tau_{\rm D}$ nearly 4 times larger in the absence of any other bright galaxies in its vicinity. This shows again that, when observing a LAE during the EoR, it is likely to be surrounded by a group of nearby bright galaxies, further emphasizing the importance of the contributions of neighboring, albeit, fainter galaxies.
	\item[\textbf{Galaxy d}.] A highly {\lya}-obscured galaxy (${\rm H}_{160}{=}27.1$ mag). This object is one order of magnitude fainter than Galaxies b and c, and has a bubble size of only ${\sim}2$cMpc. It resides in the smallest $\hii$ bubble among all bright (${\rm H}_{160}{<}27.5$ mag) galaxies and possesses a {\lya} transmission rate of only ${\sim}15$\%. This galaxy has only 2 neighbours within the mock image while both are due to random alignment.
\end{enumerate}

\section{Characterizing {\lya} transmission during the EoR}\label{sec:result}
The 4 modelled galaxies presented above demonstrate some of the characteristic {\lya} IGM transmission properties among bright LBGs during the EoR. They also clearly present the real and substantial variations in {\lya} detectability within a quite modest range of luminosities (factors of a few). In this section, we use the whole modelled high-redshift galaxy population to quantify how reionization impacts {\lya} transmission, and provide guidance for searching for ionized bubbles in the heart of the cosmic reionization epoch.

\subsection{Correlation of damping wing with the IGM properties}
At a given redshift the damping wing optical depth $\tau_{\rm D}$, on average, decreases if the LAE sits inside larger $\hii$ bubbles. This is illustrated in the upper left panel of Fig. \ref{fig:tau_R} for galaxies brighter than ${\rm H}_{160}{=}27.5$ mag. Such a $\mathcal{T}-R_{\hii}$ correlation becomes obvious when the mean density and neutral hydrogen fraction are substituted in equation (\ref{eq:tau_lya}; see the dashed line in the panel). However, as overdense regions are typically the first to become ionized, we also expect to see a positive correlation between the {\lya} transmitted fluxes and large-scale\footnote{A cubic volume of 150cMpc$^3$ is used to show the large-scale density. This choice is somewhat arbitrary with the side length, 5.3cMpc (i.e. $2'$ at $z{=}8$), corresponding to a WFC3 field of view.} overdensities. This is shown in the upper right panel of Fig. \ref{fig:tau_R} where we also observe a much larger scatter compared to the $\mathcal{T}-R_{\hii}$ relation. When focusing on currently observable galaxies (i.e. ${\rm H}_{160}{<}27.5$ mag), we see that the obscured LAEs are on average located in $\hii$ regions smaller than ${\sim}7$cMpc at $z=8$ and with an overdensity of $\Delta_{\rm 3D}{\lesssim}1$.

\subsection{Correlation of damping wing with galaxy properties}

\begin{figure*}
	\begin{minipage}{\textwidth}
		\centering
		\includegraphics[width=\textwidth]{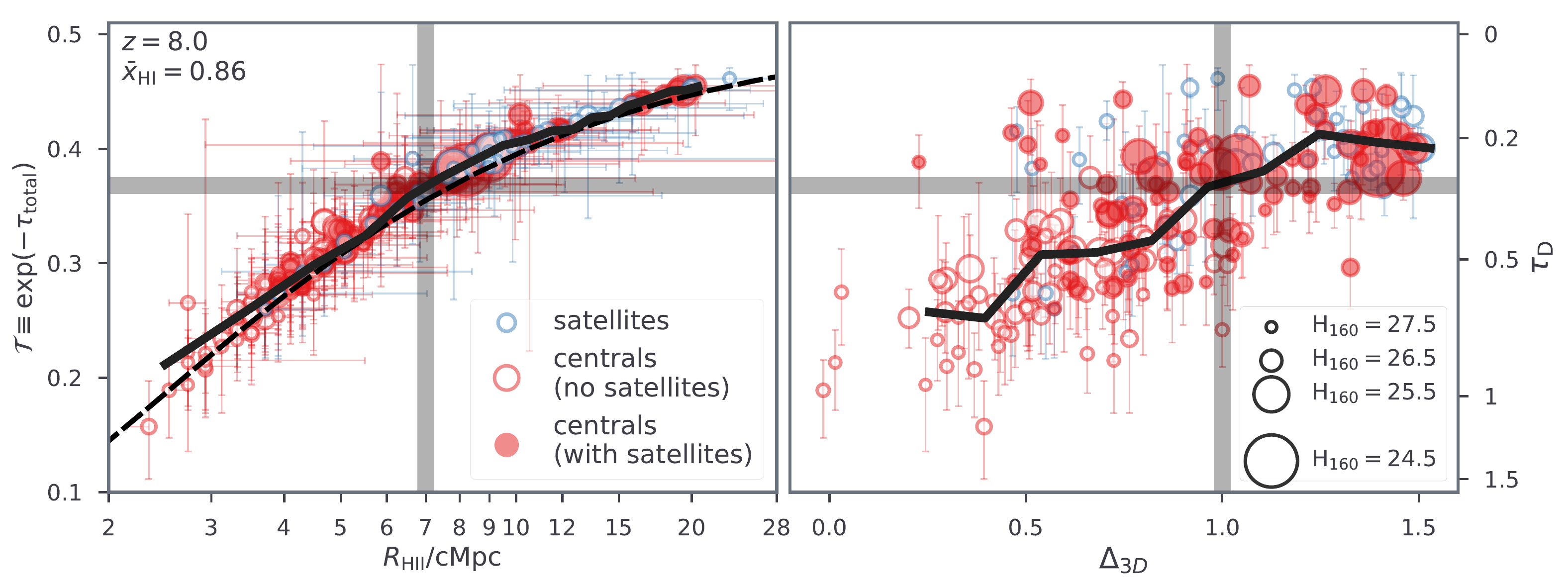}\\%\vspace*{-2.1mm}\\
		\includegraphics[width=\textwidth]{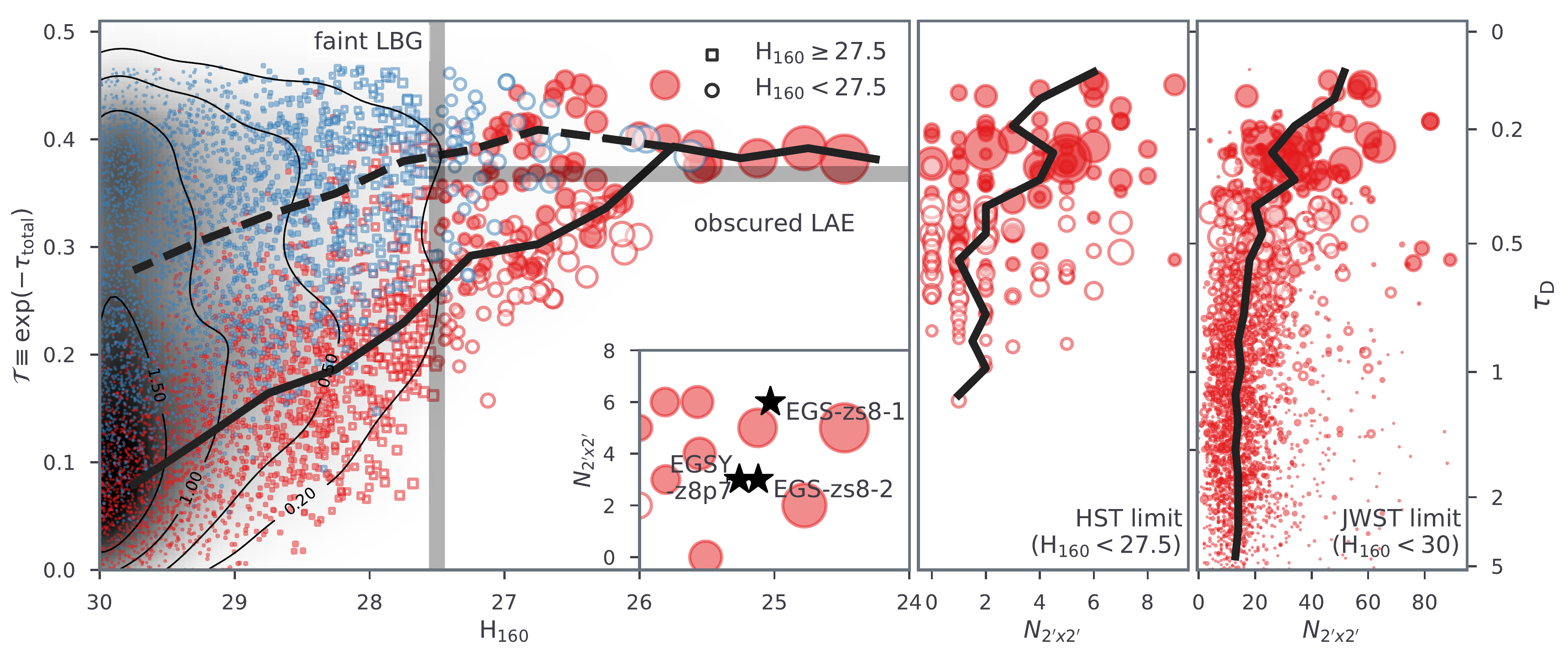}%\vspace*{-2mm}
		%\vspace*{-3mm}
		\caption{\label{fig:tau_R} {\it Upper panels:} the {\lya} transmission ($\mathcal{T}{\equiv} e^{-\tau_{\rm total}}$; left axis) or damping optical depth ($\tau_{\rm D}$; right axis) as a function of the ${\hii}$ bubble size ($R_{\rm HII}$; left panel) or overdensity smoothed over a cubic volume of $150{\rm cMpc}^3$ ($\Delta_{\rm 3D}$; right panel) for galaxies brighter than ${\rm H}_{160}{=}27.5$ mag. 
			Central galaxies (i.e. the brightest ones in their $\hii$ bubbles) and their satellites (i.e. the remaining galaxies) are shown separately using red and blue colours with increasing circle sizes representing higher luminosities.
			% and filled/empty circles indicating central galaxies with/without satellites. % (see more in the text).
			The errorbar corresponds to [16, 84] percentiles of the transmission and bubble size varying between 500 different random sightlines. 
			The median relation is indicated by the solid curve while the dashed one shows the $\mathcal{T}{-}R_{\rm HII}$ relation at the mean overdensity and neutral hydrogen fraction for comparison. $\tau_{\rm total}{=}1$ is indicated using the horizontal grey stripes with the vertical ones highlighting the corresponding $R_{\rm HII}$ and $\Delta_{\rm 3D}$ on the median relation.
			{\it Lower-left panel:} the $\mathcal{T}{-}{\rm H}_{160}$ relation with galaxies fainter than ${\rm H}_{160}{=}27.5$ mag distinguished by the square symbol and further illustrated using a 2D histogram (and contours) for better visualization. The median relations are indicated using the thick solid and dashed black lines for central and satellite galaxies, respectively. The correlation between the number of neighbouring galaxies within a WFC FoV ($2'{\times}2'$; 2D projection with $\Delta z{\sim}1$) and ${\rm H}_{160}$ is shown in the inset for the most luminous centrals. Preliminary estimates (Leonova et al., in prep.) are indicated by the star symbol for 3 objects of the $z{\sim}8$ LAEs \citep{Oesch2015ApJ...804L..30O,Roberts-Borsani2016ApJ...823..143R,Zitrin2015ApJ...810L..12Z}. {\it Lower-right two panels:} the correlation between $\mathcal{T}$ and the number of neighbours brighter than ${\rm H}_{160}{=}27.5$ or 30 mag ({\it JWST} limit). In these two panels, galaxies shown are also brighter than the {\it HST/JWST} limit, and the median relations are binned by $\mathcal{T}$. Central galaxies having satellites within their bubbles (both are brighter than the {\it HST} limit) are highlighted with filled circles.
		}
	\end{minipage}
\end{figure*}

\begin{figure*}
	\begin{minipage}{\textwidth}
		\centering
		%\hspace*{-4mm}
		\includegraphics[width=\textwidth]{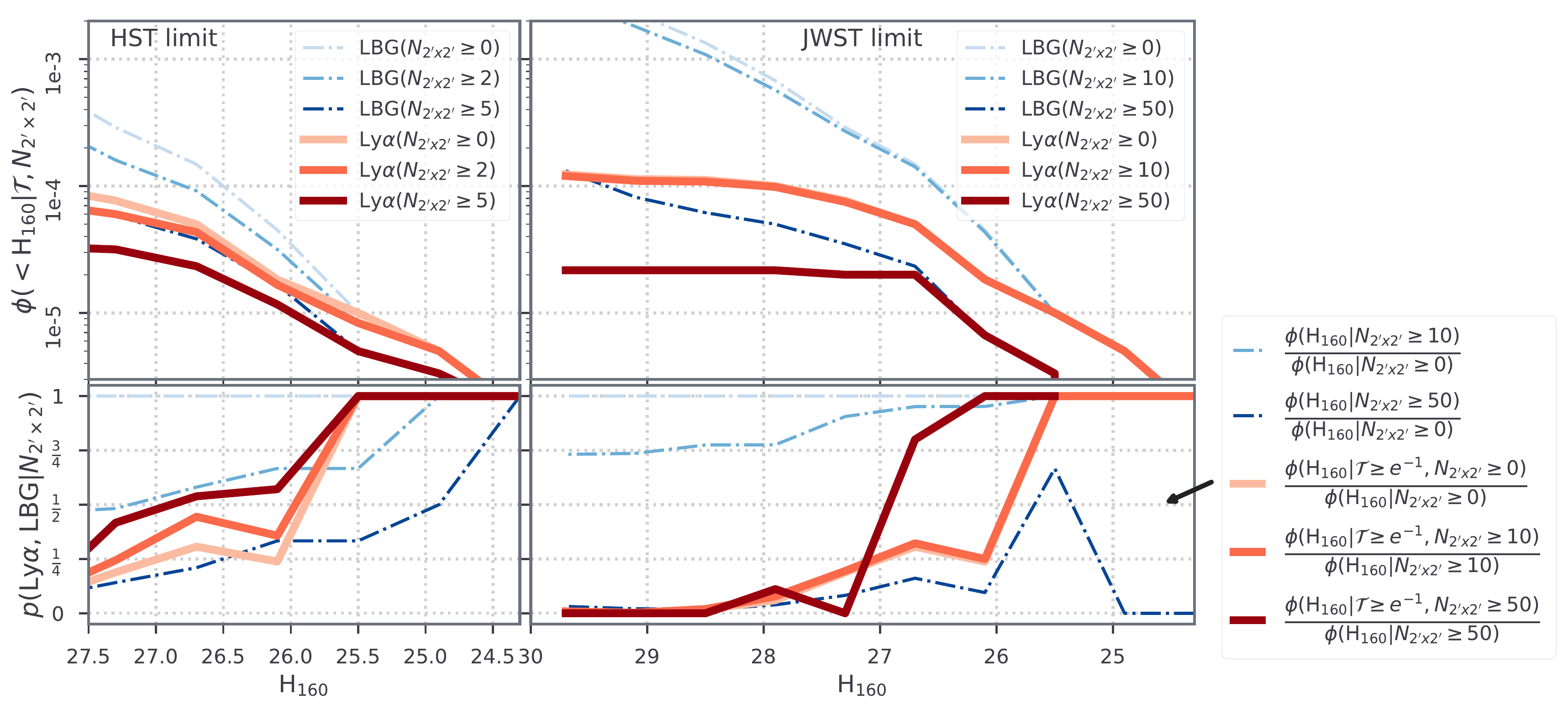}
		%\vspace*{-7mm}
		\caption{\label{fig:detectability} {{\it Upper-left panel:} the cumulative number density in units of ${{\rm cMpc^{-3} mag^{-1}}}$ as a function of galaxy UV magnitude (${\rm H}_{160}$) for all central galaxies ($\phi({<}{\rm H}_{160}|N_{2'{\times}2'}{\ge}0)$; light-blue dash-dotted thin line), centrals with more than 1 bright (i.e. ${\rm H}_{160}{<}27.5$) neighbours ($\phi({<}{\rm H}_{160}|N_{2'{\times}2'}{\ge}2)$; medium-blue dash-dotted thin line) and with at least 5 bright neighbours ($\phi({<}{{\rm H}_{160}}|N_{2'{\times}2'}{\ge}5)$; dark-blue dash-dotted thin line) as well as central galaxies with detectable {\lya} emission including all transparent LAEs ($\phi(<{\rm H}_{160}|\mathcal{T}{\ge}e^{-1}, N_{2'{\times}2'}{\ge}0)$; light-red solid thick line), and those with ${\ge}2$ (i.e. $\phi({<}{{\rm H}_{160}}|\mathcal{T}{\ge}e^{-1}, N_{2'{\times}2'}{\ge}2)$; medium-red solid thick line) or ${\ge}5$ bright neighbouring galaxies (i.e. $\phi({<}{{\rm H}_{160}}|\mathcal{T}{\ge}e^{-1}, N_{2'{\times}2'}{\ge}5)$; dark-red solid thick line). {\it Upper-right panel:} similar to the previous panel for a typical {\it HST} extragalactic field, this panel forecasts for upcoming deep {\it JWST} surveys with the detection limit for target galaxies and their neighbours updated to ${\rm H}_{160}{<}30$ and with a redshift uncertainty of $\Delta z{=}0.35$. Note that the $N_{2'{\times}2'}$ limits are changed to ${\geq}10$ (overlapped with $\phi(<{\rm H}_{160}|\mathcal{T}{\ge}e^{-1}, N_{2'{\times}2'}{\ge}0)$) and 50. 
				{\it Lower panels: } the number density of central galaxies with varying $N_{2'{\times}2'}$ normalized by $\phi({\rm H}_{160}|N_{2'{\times}2'}{\ge}0)$ (blue dash-dotted thin lines in the upper panels) as well as those with detectable {\lya} emission normalized by the total galaxy population with corresponding $N_{2'{\times}2'}$ (see the legend in the lower-right corner for {\it JWST} as an example for more details).}
			%The number density of transparent LAEs (thick red solid lines) and all central galaxies (thin blue dash-dotted lines) with a varying number of bright neighbours (i.e. $N_{2'{\times}2'}$ distinguished by the line brightness) within a WFC3 FoV ($2'{\times}2'$). For instance, all centrals, centrals with more than 1 bright neighbours (i.e. $\phi({<}{\rm H}_{160}|N_{2'{\times}2'}{\ge}2)$; thin dashed line), and those with at least 5 bright neighbours (i.e. $\phi({<}{{\rm H}_{160}}|N_{2'{\times}2'}{\ge}5)$; thin dash-dotted line)
			%While the cumulative densities are shown in the upper panels, the ones normalized by the number density of all central galaxies (i.e. the blue dash-dotted lines in the upper panel) with the corresponding $N_{2'{\times}2'}$ are presented in the bottom panels. 
			%The difference between the left and right panels is the detection limit -- both the target galaxies and their neighbours (left for a typical HST extragalactic field, right for a deep JWST survey).}
		}
	\end{minipage}
\end{figure*}

The lower-left panel of Fig. \ref{fig:tau_R} shows the relation between galaxy UV luminosity and {\lya} transmission. In general, brighter galaxies possess higher {\lya} transmission as their UV ionizing luminosities are more effective at ionizing the surrounding IGM and therefore creating larger $\hii$ bubbles \citep{Geil2017MNRAS.472.1324G,Davies2021MNRAS.501..146D}. However, we also see a bimodal distribution in the $\mathcal{T}-{\rm H}_{160}$ relation. After separating galaxies according to their luminosities in each $\hii$ bubble, we find that the two peaks originate from different galaxy populations. Central galaxies, defined as the brightest galaxy in each bubble, dominate the local ionizing photon budget and therefore have lower $\tau_{\rm D}$ with increasing luminosities. However, most galaxies brighter than ${\rm H}_{160}{\sim}25.5$ mag have a similar level of damping-wing absorption ($\tau_{\rm D}{\sim}0.27$ or $\mathcal{T}{\sim}0.38$) and all of them are transparent in {\lya}. This is consistent with the 4 observed bright LBGs showing strong {\lya} emission \citep{Oesch2015ApJ...804L..30O,Zitrin2015ApJ...810L..12Z,Roberts-Borsani2016ApJ...823..143R,Stark2017MNRAS.464..469S} and suggests that the brightest, color-selected LBGs are likely to reside in large $\hii$ regions (see estimate of the bubble size by \citealt{Tilvi2020ApJ...891L..10T} for one of the LAEs) as predicted by generic inside-out models of reionization.
%promising candidates to follow up with spectroscopy in order to find their {\lya} emissions. 
On the other hand, fainter satellite galaxies have their $\hii$ bubble size mostly set by the central galaxies. Therefore, we see a horizontal offset towards lower luminosity\footnote{In other words, satellite galaxies (by definition) are fainter than their centrals in an ionized region with a given optical depth.} from the median $\mathcal{T}-{\rm H}_{160}$ relation of central galaxies (shown as the solid curve) to that of the satellites (dashed curve). Comparing between the two galaxy populations, we see that galaxies fainter than ${\rm H}_{160}{\sim}27.5$ can only be detected in {\lya} when they share the same {$\hii$} bubble as another brighter one.

We showed in Section \ref{subsubsec:showcases} that, despite being more than one magnitude fainter than the brightest galaxy in our simulation (i.e. Galaxy a), Galaxy b possesses a much higher {\lya} transmission rate as its high overdensity environment leads to a great number of bright galaxies contributing to the local ionizing photon budget, which pushes the ionizing front much further. Looking at the $\mathcal{T}-\Delta_{\rm 3D}$ panel of Fig. \ref{fig:tau_R}, we see that, compared to those dominating their UV background alone (empty red circles), galaxies with neighbours in their $\hii$ regions (filled red circles) are more likely to reside in high-density regions, and are less obscured by the {\lya} damping-wing absorption.

To illustrate this in more detail,
we select mock (central) galaxies brighter than ${\rm H}_{160}{=}27.5$ mag and plot their {\lya} transmission fluxes as a function of the number of neighbours (also with ${\rm H}_{160}{<}$27.5 mag) in the second panel on the bottom of Fig. \ref{fig:tau_R}. As before, this luminosity threshold is chosen to match the current observational limit (e.g. CANDELS DEEP $5\sigma$ depth, see e.g. \citealt{Roberts-Borsani2016ApJ...823..143R}) and the number count is done in a $2'{\times}2'$ FoV centred at the target object with an observational depth of $\Delta z{\sim}1$. We see a positive correlation between $\mathcal{T}$ and $N_{2'{\times}2'}$ -- galaxies with an increasing number of neighbours are more likely to have their surrounding gas ionized to a great distance, and therefore their {\lya} emission is less obscured by damping-wing absorption. {It is worth noting that, as transparent LAEs are located in bubbles with a radius of ${\gtrsim}7$cMpc (larger than the FoV in the mock image), their neighbouring galaxies are also likely to be observable in {\lya}. From the second bottom panel of Fig. \ref{fig:tau_R}, we see that, on average, ${\gtrsim}4$ LAEs can be found in those highly ionized regions.}

\subsection{Searching for ionizing bubbles at $z{\ge}7$}

The lower-left panel of Fig. \ref{fig:tau_R} illustrates that the majority of modelled galaxies with ${\rm H}_{160}{\lesssim}26$ mag have bright neighbours within a WFC3 FoV. This agrees with preliminary results (Leonova et al., in prep; see also \citealt{Tilvi2020ApJ...891L..10T}) from an environmental study of 3 confirmed $z{\sim}8$ LAEs (i.e. EGS-zs8-1, EGS-zs8-2 and EGSY-z8p7; \citealt{Oesch2015ApJ...804L..30O,Roberts-Borsani2016ApJ...823..143R,Zitrin2015ApJ...810L..12Z}) -- up to 6 LBGs brighter than ${\rm H}_{160}{\sim}27.5$ mag were revealed in each of their ${\it HST}$ WFC3 images (see the star symbols in Fig. \ref{fig:tau_R}). {Finding more of these objects will help understand their properties and impact during the EoR, as well as to infer the ionizing bubble size and to quantify the morphology of reionization.} Therefore, in this subsection, we further explore the strategy to search for LAEs during the EoR. 

In the upper-left panel of Fig. \ref{fig:detectability}, we present the cumulative number density, $\phi({<}{\rm H}_{160})$, for all central galaxies with a varying number of bright neighbours as well as for those classified as transparent LAEs. The fraction of modelled LBGs with detectable {\lya} emission, $p({\rm Ly\alpha})$, is an informative quantity. Not only can it place constraints on the neutral hydrogen fraction (e.g. \citealt{Mesinger2015MNRAS.446..566M,Mason2018ApJ...856....2M}), in a blind survey of {\lya} emission among known high-redshift LBGs, $p({\rm Ly\alpha})$ is also an indicator of the detection rate. The lower left panel of Fig. \ref{fig:detectability} presents $p({\rm Ly\alpha})$ as a function of the galaxy UV luminosity and $N_{2'\times2'}$ together with the fraction of LBGs having different number of neighbours for comparison.

We find that the success rate for finding LAEs among LBGs regardless of their environment is 100\% when targeting the brightest galaxies (i.e. ${\rm H}_{160}{\lesssim}25.5$ mag). This rate drops significantly towards the fainter end (e.g. $20-30$\% at ${\rm H}_{160}$ between 26 and 27.5 mag). Considering only LBGs surrounded by bright neighbours can increase the detectability. For instance, at the current observing limit (${\rm H}_{160}{<}27.5$ mag), we find that following up galaxies with at least 2 neighbours can boost the detectability up to 40\%. This number rises to nearly 60\% when focusing on galaxies with ${\ge}5$ neighbours. However, as shown by the LBG fraction curves in Fig. \ref{fig:detectability}, only ${\lesssim}$ 25\% of galaxies brighter than ${\rm H}_{160}{=}26$ mag have $N_{2'{\times}2'}{\ge}5$ neighbours. Therefore, we conclude an optimal survey strategy for finding LAEs is to follow up with spectroscopy all LBGs brighter than ${\rm H}_{160}{=}26$ mag while only focusing on fainter ones with at least 2 neighbours.

In the coming decade, {\it JWST} will allow study of galaxies at magnitudes much fainter than ${\rm H}_{160}{=}27.5$ mag (i.e., the empty squares shown in Fig. \ref{fig:tau_R}). To model this we reconsider the detection threshold to be ${\rm H}_{160}{=}30$ mag, and search for neighbouring galaxies that will be detectable by {\it JWST}. For simplicity, we use the same FoV\footnote{With NIRCam, {\it JWST} observes two areas of $2.2'{\times}2.2'$ FoV separated by a gap of ${\sim}0.7'$.} as the mock {\it HST} WFC3 survey but assume reduced photometric uncertainties and consider $\Delta z=0.35$ (i.e. our simulation box size). From the lower right panel of Fig. \ref{fig:tau_R}, we see the number of neighbours increases by a factor of ${\sim}10$ and that, on average, ${\gtrsim}25$ galaxies (which are also observable in {\lya}) with ${\rm H}_{160}{<}30$ mag are found in the area surrounding transparent LAEs. Consequently, we will need to target galaxies with many more neighbours to see an influence of neighbours on $\tau_{\rm D}$, since $N_{2'{\times}2'}{\ge}10$ is nearly identical\footnote{{The {\it JWST} NIRCam grism programs (e.g. \citealt{Malkan2021jwst.prop.1571M,Naidu2021jwst.prop.2279N,Oesch2021jwst.prop.1895O}) will obtain 3D information and therefore the spatial correlation will become much improved from better redshift resolution. However, in this case, even with our reduced redshift uncertainties ($\Delta z{=}0.35$), we may be still overestimating the foreground/background sources.}} to $N_{2'{\times}2'}{\ge}0$ (i.e. all transparent LAEs). From the right panels of Fig. \ref{fig:detectability}, we find a number of ${\sim}$50 neighbours is optimal in order to reach a detection rate of ${>}50$\% at ${\rm H}_{160}{\lesssim}27$ mag. At even fainter magnitudes, the {\lya} emission from most galaxies will be highly attenuated by damping-wing absorption.
	 
\section{Conclusion}\label{sec:conclusion}
In this work, we study {\lya} transmission in the IGM by forward-modelling the damping-wing absorption using {\meraxes}, a coupled galaxy formation and reionization SAM that is consistent with the measured high-redshift LBG luminosity functions and EoR history inferred from {\it Planck} \citep{Qin2017a}. As more luminous galaxies (and their neighbours) are able to ionize surrounding inter-galactic hydrogen to a much larger radius than fainter galaxies, we find the inter-galactic {\lya} absorption becomes weaker towards brighter galaxies. We assume the detectability of {\lya} to be determined by the damping-wing optical depth (hence the size of the ionized bubble in which they sit) and define LBGs with a total {\lya} optical depth less than 1 to be transparent LAEs. Our model shows that transparent LAEs will be found in {$\hii$} bubbles that are at least 7cMpc in size at $z=8$ (where the global neutral hydrogen fraction is predicted to be ${\sim}86$\%). 

In particular, we find that all modelled LBGs with ${\rm H}_{160}{\lesssim}25.5$ mag are detectable LAEs. This is consistent with the high detection rate (i.e. $p({\rm Ly}\alpha){=}100$\%) of {\lya} emission in recent spectroscopic follow-up of the 4 brightest galaxies at $z{\sim}8$ \citep{Oesch2015ApJ...804L..30O,Zitrin2015ApJ...810L..12Z,Roberts-Borsani2016ApJ...823..143R,Stark2017MNRAS.464..469S,Tilvi2020ApJ...891L..10T}. These results, in addition to the more recent findings of $z{\sim}8$ LAEs by \citet{Jung2020arXiv200910092J,Endsley2020arXiv201003566E} and \citet{Laporte2021arXiv210408168L} indicate very high LAE fractions among luminous galaxies beyond $z{\sim}6$. This is in contrast to measurements of fainter galaxies in deep fields or using less-massive lensed galaxies, which show a significant drop of $p({\rm Ly}\alpha)$ from $z=6$ to 8 \citep{Schenker2014ApJ...795...20S,Hoag2019ApJ...878...12H}. Our simulation (see Fig. \ref{fig:snapshot}) suggests that the combination of these measurements at both high and low luminosity provides evidence for massive LAEs being likely to reside in large {$\hii$} bubbles.

We also find that reionization is more advanced around galaxies in high-density regions compared to those that are isolated, with galaxies having a larger number of bright neighbours being more likely to reside in large $\hii$ bubbles, leading to {\lya} emission that is less attenuated by the IGM. This not only provides evidence that overdensity plays an important role in driving reionization, but can also motivate searching for LAEs in high-density environments during reionization. Transparent LAEs, on average, are located in overdense regions and are found to possess ${\gtrsim}2(25)$ neighbouring galaxies brighter than ${\rm H}_{160}{\sim}27.5(30)$ mag within mock images of $2'{\times}2'$ FoV. As their neighbours are also within the same ${\hii}$ bubbles, they are also likely to be observable in {\lya}.

{Finding more high-redshift LAEs will help quantify the morphology of the EoR and infer the properties of galaxies responsible for driving the reionization. We find that while nearly 70--80\% of $z{=}8$ galaxies with ${\rm H}_{160}$ between 26 and 27.5 mag experience strong damping-wing absorption, targeting those surrounded by bright neighbours can significantly increase the incidence of {\lya} emission. For example, at the current observational limit, 40\% galaxies with more than 1 neighbour would be considered detectable LAEs. This number increases to 60\% when focusing on galaxies with at least 5 neighbours. Finally, we predict that upcoming {\it JWST} observations are likely to reveal a factor of 10 more neighbouring galaxies.  With such a large sample size, we find that targeting galaxies with ${\sim}50$ neighbours will yield a success rate of more than 50\% for finding LAEs among LBGs during the EoR. These large samples from JWST will provide insights into the morphology and scale sizes of the ionized regions, and their development and growth with redshift through the EoR, that is just not possible with {\it HST}.

\section*{Acknowledgements}
This research 
was supported by the Australian Research Council Centre of Excellence for All Sky 
Astrophysics in 3 Dimensions (ASTRO 3D), through project \#CE170100013. Parts of this work was supported by the European Research Council (ERC) under the European Union’s 
Horizon 2020 research and innovation programme (AIDA -- \#638809). The results presented here reflect the authors’ views; the ERC is not responsible for their use. The simulations presented in this work were run on the OzSTAR national facility at Swinburne University of Technology. PAO acknowledges support from the Swiss National Science Foundation through the SNSF Professorship grant 190079. The Cosmic Dawn Center (DAWN) is funded by the Danish National Research Foundation under grant No.\ 140. GDI is grateful for support from program HST-GO-15103.002-A from STScI/AURA Inc under NASA contract NAS 5-26555. RPN gratefully acknowledges an Ashford Fellowship granted by Harvard University.

\section*{Data availability}
The data underlying this article will be shared on reasonable request to the corresponding author.

\bibliographystyle{\dir mn2e}
\bibliography{reference}

%\appendix

%\bsp
\label{lastpage}
\end{document}